# Preparation and electrochemical properties of nitrogen-doped starch hard carbon anode materials for lithium-ion battery

Aoqi Huang *, Yibo Tu, Qichao Yu

School of Electronics and Information Engineering, Hangzhou Dianzi University, Hangzhou 310018, China;

* Corresponding author.
  E-mail address: haq@hdu.edu.cn(A. Huang)

**Abstract:** Here, we report the synthesis of hard carbon materials(CSH) made from corn starch and their application as an anode in lithium-ion batteries. The study shows that the Microstructure and electrochemical properties of CSHs are affected by nitrogen doping. It is found that nitrogen is embedded in the carbon layer with graphite nitrogen, pyridine nitrogen, and pyrrole nitrogen, so as to the surface morphology was changed and reduced the disorder of the materials. The electrochemical test results show that the introduction of nitrogen elements can increase the reversible capacity of the material, with the first discharge capacity reaching above 426.35 mAh g-1, and the rate performance also improves. When triethylenetetramine and pre-carbonized corn starch are carbonized at a mass ratio of 1:9, the obtained material has a reversible capacity of 122.04 mAh g-1 at a rate of 2 C. During the carbonization process, the nitrogen in triethylenetetramine is doped into the carbon materials, improving the electrochemical performance of the material.

## 1. Introduction

As a crucial component of battery systems, negative electrode materials significantly impact battery performance. Negative electrode materials can be divided into carbon-based and non-carbon-based materials. Non-carbon-based materials have higher theoretical capacities, as exemplified by metal-based negative electrode materials. Lithium ions react chemically with active metals in these materials to form $Li_xM$ compounds, where M is the metal and $x > 1$. This is one of the reasons why the capacity of intermetallic compounds is significantly higher than that of carbon-based negative electrode materials [1]. However, the enormous volume change limits the further development of non-carbon-based negative electrodes, such as the volume change rate of silicon-based negative electrodes reaching 400% [2]. Among carbon-based negative electrode materials, graphite is a successful commercial negative electrode material with a high theoretical capacity. However, due to its well-ordered layered structure, it is difficult to perform large-scale charge and discharge, and graphite often needs to be modified to obtain better electrochemical performance.

To obtain high-quality negative electrodes with faster charging speeds and lower costs, hard carbon materials have emerged in the sights of researchers. Hard carbon, as a material that is difficult to graphitize, features short-range order and long-range disorder of graphene layers randomly distributed and is also known as non-graphitizable carbon materials [3]. This material has a rich microporous structure [4] and microporous lithium storage method [5][6][7], providing numerous lithium storage sites, and this structure results in almost no volume change during the battery cycling process [8], making it one of the ideal negative electrode materials.

Various plants, with their unique hierarchical porous structures, ensure effective cycling and non-organic exchange, allowing for growth and the continuation of life in the natural world [9][10]. These advantages mean that carbonized biomass materials have diverse microscopic structures and exhibit different characteristics for lithium-ion storage. Due to their low cost, renewability, and environmental friendliness, biomass materials have broad application prospects in negative electrode materials. Starch, as a typical representative of biomass, has the same chemical formula $(C_6H_{10}O_5)_n$ for various starches. The presence of van der Waals forces and intermolecular hydrogen bonds further promotes the formation of complex starch molecular structures. Based on the different glycosidic linkage modes, the molecular structure of starch can be divided into amylose and amylopectin. Starch's unique multicrystalline structure is related to these two parts, with the crystalline region usually composed of amylopectin and the amorphous region composed of amylose [11][12]. The content of amylopectin in starch is about 80%, and amylopectin has a spatial structure similar to graphite, making it a potential high-capacity negative electrode material for lithium-ion batteries. To obtain higher-performance hard carbon negative electrodes, researchers often modify negative electrode materials through activation and doping with heteroatoms [13][14]. Introducing heteroatoms into the crystal lattice can bring positive properties to carbon materials, such as improving electrical conductivity and introducing different sites and functional groups for lithium-ion adsorption [15][16]. Since melamine has rich nitrogen content, this paper uses melamine to dope corn starch to obtain hard carbon materials with better electrochemical properties and improved negative electrode materials.

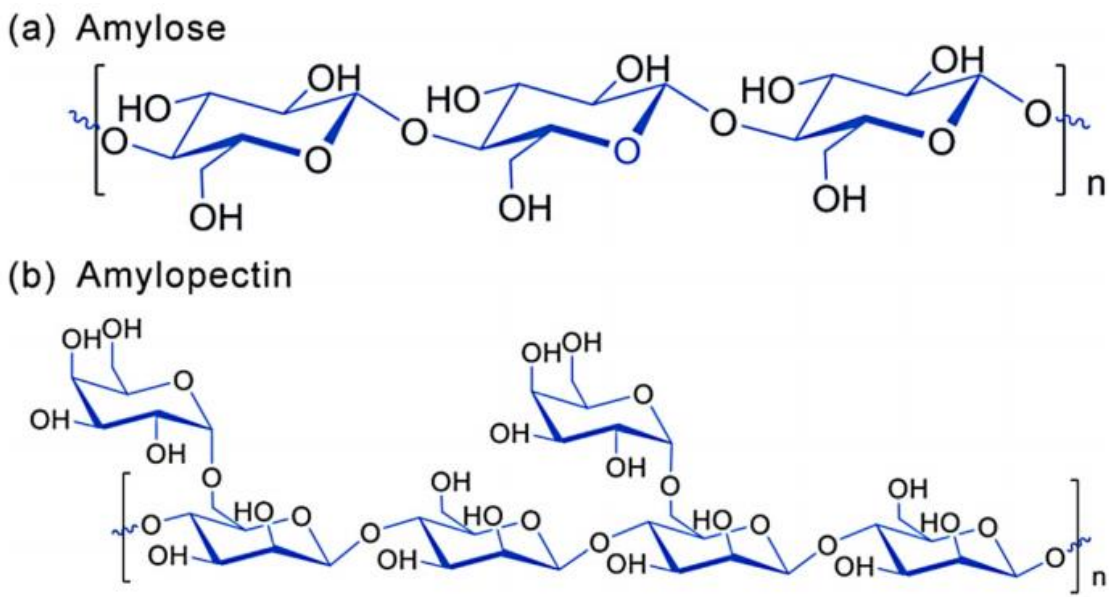

Figure 1. Structural illustration of starch. The monomeric units and typical functional groups of (a) amylose parts [15]and (b) amylopectin parts within starch[16].

## 2. Experimental
### 2.1. Reagents and Instruments

The main raw materials are corn starch (Food grade); Melamine (analytically pure), Tianjin Huasheng; Hydrochloric acid, Shanghai Macklin Biochemical Technology Co., Ltd.

The surface topography of the samples was observed using scanning electron microscopy (SEM, Sigma 300, ZEISS, Germany), the phase characterization of the samples was performed by X-ray diffractometer (XRD, D2 Phaser, Bruker, Germany), the surface defects of the materials were analyzed by Raman spectroscopy (Raman, LabRam HR Evolution, Horiba, Japan), and X-ray photoelectron spectroscopy (XPS, Scientific K-Alpha, Micromeritics, USA) analyzes the surface elements of materials.

The charge and discharge test of the battery is carried out on the Xinwei battery testing system, and the electrochemical impedance test is carried out on the electro-chemical impedance test produced by ZAHNER company in Germany.

### 2.2 Sample preparation

The obtained hard carbon materials are named CSH, NCSH1, NCSH2, and NCSH3.

In this process, corn starch is first placed in a muffle furnace under a nitrogen atmosphere and pre-carbonized at 400°C. Next, the pre-carbonized material is thoroughly milled and then sieved through a 325-mesh sieve. The sieved material is then placed in 1M HCl solution and stirred for 5 hours. After that, the material is filtered out, washed repeatedly with deionized water until the filtrate pH is neutral, and finally dried in a drying box at 80°C to obtain pre-treated starch hard carbon HC.

The preparation of CSH involves filling HC into a graphite crucible and carbonizing it under a nitrogen atmosphere. With a heating rate of 5°C/min, the temperature is raised to 1000°C and maintained for 2 hours for carbonization treatment. The temperature is then allowed to cool naturally to room temperature, and the resulting CSH hard carbon material is obtained by removing the crucible.

The preparation of NCSH involves mixing HC with melamine in different mass ratios (9:1, 8:2, 7:3). The mixture is thoroughly milled in a mortar until the carbon material and melamine are evenly mixed. The mixed material is then carbonized in a tubular furnace at 1000°C for 2 hours, with a heating rate of 5°C min-1 and a nitrogen atmosphere. After the material has naturally cooled, it is taken out and prepared for electrode assembly. The samples are named according to the mass ratio of melamine as NCSH1, NCSH2, and NCSH3.

### 2.3. Preparation of electrode sheets

The preparation of electrodes involves mixing the active material, graphite, and binder (PVDF) in a mass ratio of 8:1:1. N-methyl-2-pyrrolidone (NMP) is used as the solvent, and the mixed material is evenly coated onto a smooth and flat copper foil. The coated material is then dried in a vacuum oven at 80°C for 12 hours, followed by cutting into 12 mm electrode pieces with accurate weight. A metallic lithium foil is used as the counter electrode, Celgard as the separator, and a 1 mol/L LiPF6/[ethylene carbonate (EC) + diethyl carbonate (DEC) + ethylene carbonate monoethyl ether (EMC)] (volume ratio 1:1:1) mixed solvent as the electrolyte. The cells are assembled in a glove box filled with argon and allowed to stand for 12 hours before the next step of electrochemical performance testing.

## 3. Results and discussion
### 3.2. Morphological analysis of materials

The SEM images of CSH, NCSH1, NCSH2, and NCSH3 are shown in Fig. 2. After milling and carbonization, all materials exhibit irregular shapes and smooth, flat surfaces, with particle sizes ranging around 40 μm after passing through a 325 mesh sieve. As the melamine content increases, 5 μm pores appear on the sample surfaces, which are formed by etching during the high-temperature carbonization process, indicating that the melamine addition has a certain impact on the sample morphology. Fig. 2e shows the EDS image of NCSH1, where C elements serve as the base, and N and O elements are uniformly distributed on the sample surface, indicating that N elements have been successfully and uniformly incorporated into the carbon material during the high-temperature carbonization process.

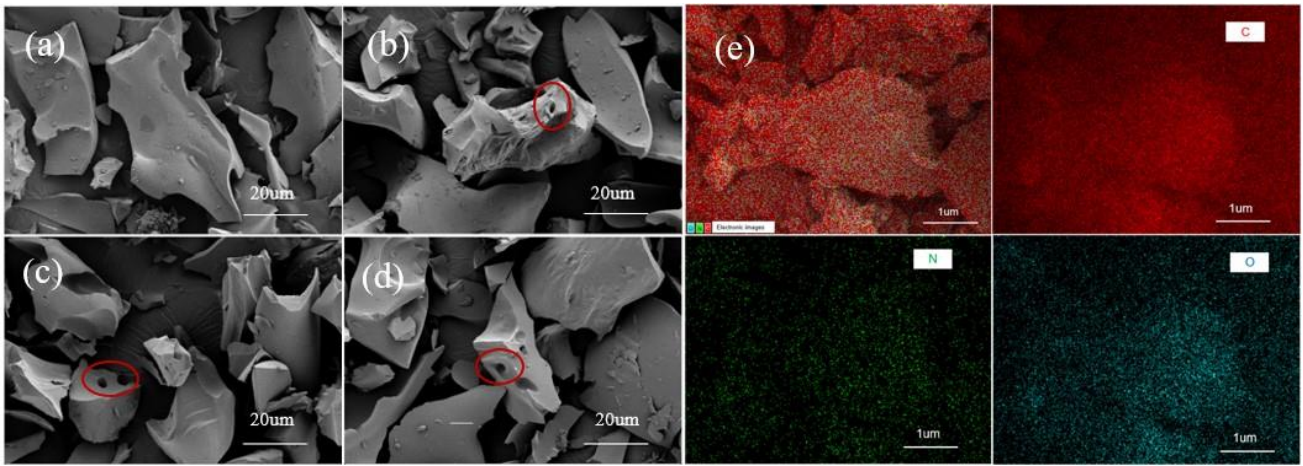

Figure 2. SEM images of starch hard carbon: (a) CSH; (b) NCSH1; (c) NCSH2; (d) NCSH3; (e) EDS images of NCSH1

### 3.2. Phase analysis of materials

XRD analysis was performed on the materials, and the results are shown in Fig. 3. The presence of two distinct broad peaks at approximately 2θ values of 23° and 43° corresponds to the (002) and (100) crystal planes of the hard carbon material, which are characteristic diffraction peaks of hard carbon materials, indicating poor crystallinity and the amorphous nature of the material. Melamine doping does not change the hard carbon material derived from starch into non-layered graphite. By comparing the peak positions, it was found that the (002) crystal plane of the melamine-doped hard carbon material shifted to the right, indicating that melamine doping reduces the interlayer spacing of the material, but the interlayer spacing is still significantly larger than that of graphite (0.34 nm). The larger interlayer spacing in hard carbon materials is beneficial for the movement of lithium ions in the negative electrode.

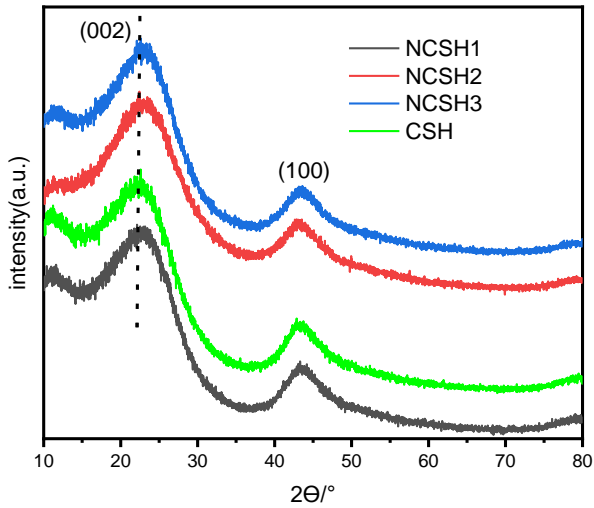

Figure 3. XRD Patterns for CSH, NCSH1, NCSH2, NCSH3

Raman spectroscopy was performed on the materials to understand their graphitization degree and defect level, and the resulting Raman spectra are shown in Fig. 4. The characteristic D and G bands of the carbon lattice appear near the incident light wavenumbers of 1356 and 1590 cm-1, respectively. The intensity of the D band is related to defects on the material surface, while the intensity of the G band is associated with the vibration of graphitic carbon atoms. The ratio of D band to G band intensity (Id/Ig) can quantify the graphitization degree of the hard carbon material and indicate the defect level, with a higher ratio indicating a higher defect density. As the melamine doping ratio increases, the Id/Ig values gradually decrease, indicating that melamine doping can change the surface defects of starch-derived hard carbon at high temperatures, reduce the material's disorder, and increase the graphitization degree. Furthermore, melamine doping helps to reduce the defect density in the materials.

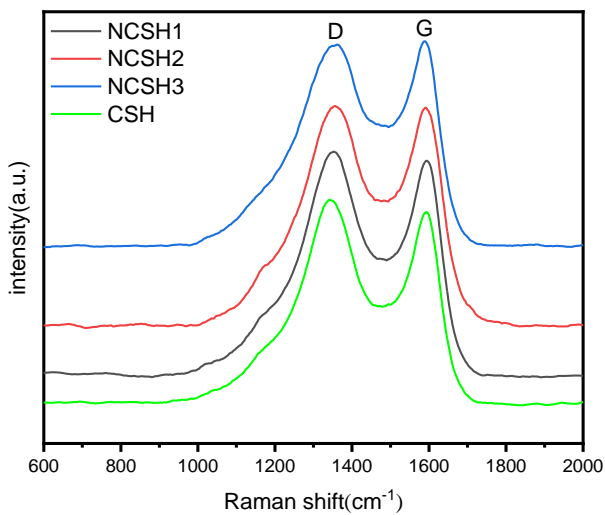

Figure 4. Raman spectra for CSH, NCSH1, NCSH2, NCSH3

XPS testing was conducted on NCSH materials to determine the existence of elements in their forms. The XPS spectrum results of the materials are shown in Fig. 4. Fig. 5(a) shows the XPS survey spectra of the three materials, where it can be found that the materials contain C, N, and O elements, and the N1s peak becomes stronger as the triethylenetetramine ratio increases. By fitting the C1s spectrum, two peaks were obtained, as shown in Figure 5b, with binding energies of 286.58 eV and 284.80 eV, corresponding to C-O and C-C chemical bonds. By fitting the N1s spectrum, several characteristic peaks were obtained, as shown in Fig. 5(c), with binding energies of 398.3 eV, 400.01 eV, and 401.1 eV, and it was determined through binding energy analysis that the nitrogen was embedded in pyridinic N, pyrrolic N, and graphitic N in amorphous carbon. It can be found that all three types of nitrogen are successfully incorporated into NCSH2, while only graphitic N and pyrrolic N are present in NCSH1, and only graphitic N and pyridinic N are present in NCSH3. As the triethylenetetramine ratio increases, the peak area of graphitic N increases significantly, and different types of nitrogen have different effects on the electrochemical performance of the materials. Finally, the structural parameters of different materials are summarized, as shown in Table 1.

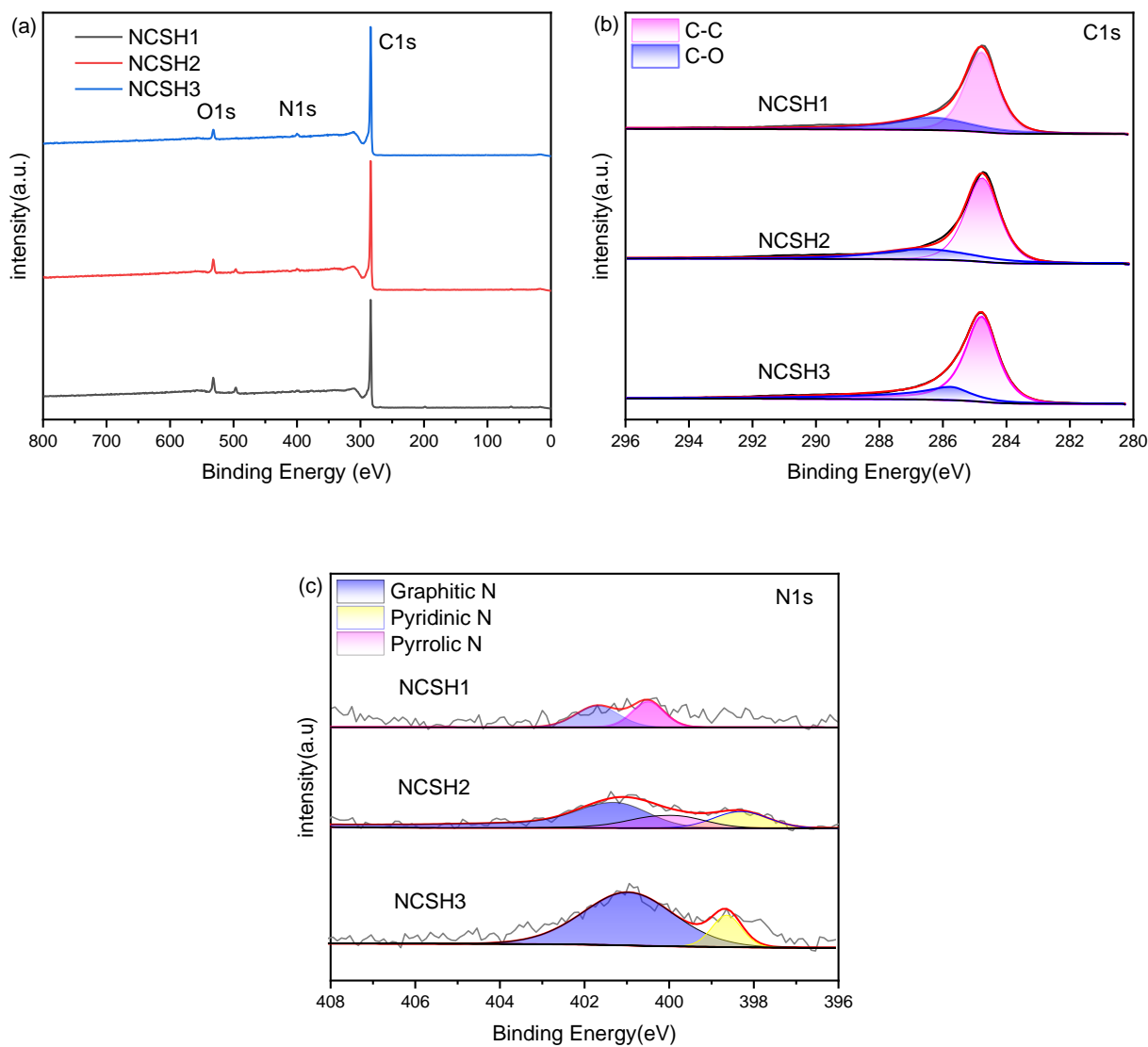

Figure 5. (a) XPS Full Spectrum; XPS fine spectra:(b)C1s;(c)N1s

Table 1. Structural parameters of different materials

| Sample | 2Θ | d(002)(nm) | $I_d/I_g$ | C、N、O Elemental content | | |
|---|---|---|---|---|---|---|
| | | | | C(%) | N(%) | O(%) |
| CSH | 22.492 | 3.9496 | 1.06 | - | - | - |
| NCSH1 | 22.704 | 3.9133 | 1.03 | 90.82 | 0.91 | 8.28 |
| NCSH2 | 22.863 | 3.8865 | 1.01 | 92.54 | 1.06 | 6.41 |
| NCSH3 | 22.757 | 3.9043 | 0.98 | 93.15 | 1.59 | 5.26 |

**2.3 Electrochemical performance analysis**

The obtained materials were assembled into a button half-cell for electrochemical performance testing. The electrochemical performance of corn starch hard carbon electrodes was studied, with the first discharge curve of the materials at a current density of 0.2 C shown in Fig. 6. The first charge capacities of NCSH1, NCSH2, NCSH3, and CSH are 248.81 mAh g-1, 249.41 mAh g-1, 238.96 mAh g-1, and 236.28 mAh g-1, respectively, and the discharge capacities are 428.01 mAh g-1, 426.35 mAh g-1, 441.37 mAh g-1, and 372.9 mAh g-1, respectively. The first cycle Coulombic efficiencies are 58.13%, 58.5%, 54.14%, and 63.36%, respectively. It can be found that the discharge capacity of the materials increases significantly after nitrogen doping, which is due to the increase in active sites caused by the nitrogen doping process, increasing lithium storage sites. Nitrogen doping can effectively increase the discharge capacity, but during the first cycle, it consumes more lithium ions, leading to a decrease in Coulombic efficiency.

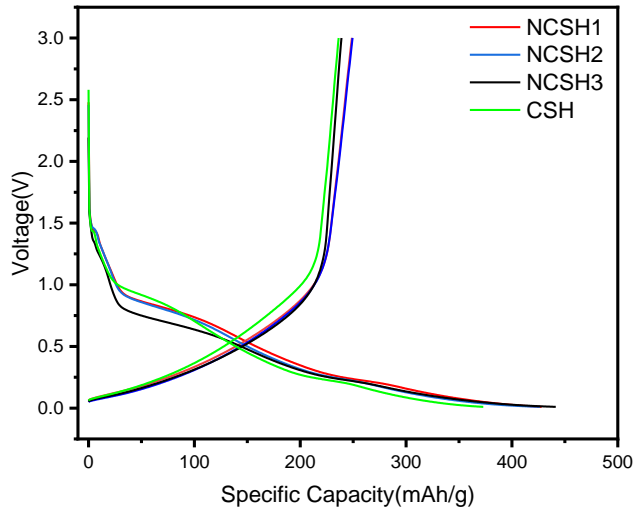

Figure 6. First-week charge-discharge curves for different materials

At a current density of 0.2 C, 100 cycles of charging and discharging were performed on the nitrogen-doped hard carbon materials, and the experimental results are shown in Fig. 7. The obtained materials all exhibit excellent cycle performance. As the number of cycles continues to increase, NCSH1 shows relatively high reversible capacity, with a capacity of 273.65 mAh g-1 after 100 cycles. The Coulombic efficiency of all materials approaches 100% when the number of cycles reaches 100, indicating excellent cycle performance. As the cycles progress, the reversible capacity of the materials shows an upward trend, which can be attributed to the involvement of active nitrogen in the electrochemical reaction, leading to an increase in reversible capacity.

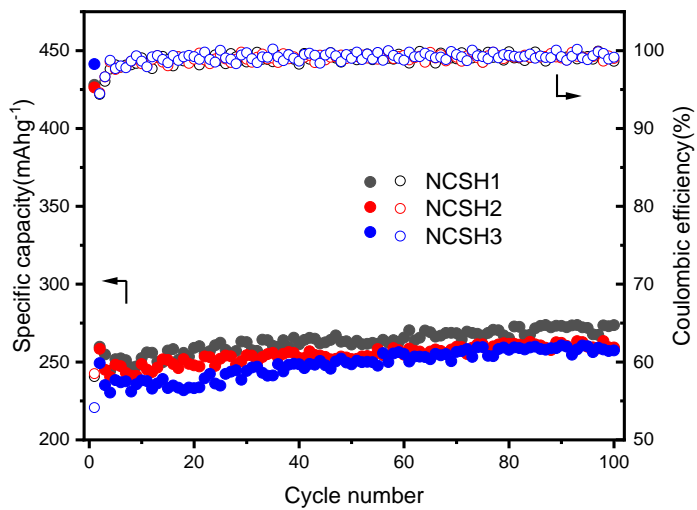

Figure 7. Cycle performance of different materials

Fig. 8(a) shows the schematic diagram of the rate performance of the hard carbon electrode. After performing charging and discharging tests at 0.2 C, 0.5 C, 1 C, 2 C, and then returning to 0.2 C, it can be found that after nitrogen doping, the capacities of the materials are higher than those without nitrogen doping at the first three rates. When the current density continues to increase to 2 C, NCSH1 and NCSH2 show relatively good rate performance, with reversible capacities of over 120 mAh g-1 at a current density of 2 C. When the current density is restored to 0.2 C after the high rate cycle, the capacities of the nitrogen-doped materials recover to a higher level. The introduction of nitrogen elements into the carbon materials forms more organic groups and lithium storage sites, providing better pathways for lithium-ion movement during the lithium-ion charging and discharging process, thereby improving the rate performance of the materials. To further understand the charging and discharging situation of the materials during the rate cycle, the rate charging and discharging curves of NCSH1 are shown in Fig. 6(b). It can be found that in the rate cycle process, the material has a large platform capacity when the current density is 0.2 C, and as the current density increases, the capacity in the platform area (0-0.1 V) gradually decreases, which corresponds to the intercalation reaction of lithium ions. It can be found that corn starch hard carbon materials mainly store lithium ions in the slope area (0.1-3 V) due to the poor crystallinity and abundant defective pores of hard carbon materials, and the slope area capacity corresponds to the adsorption of lithium ions on the surface defects of the carbon materials.

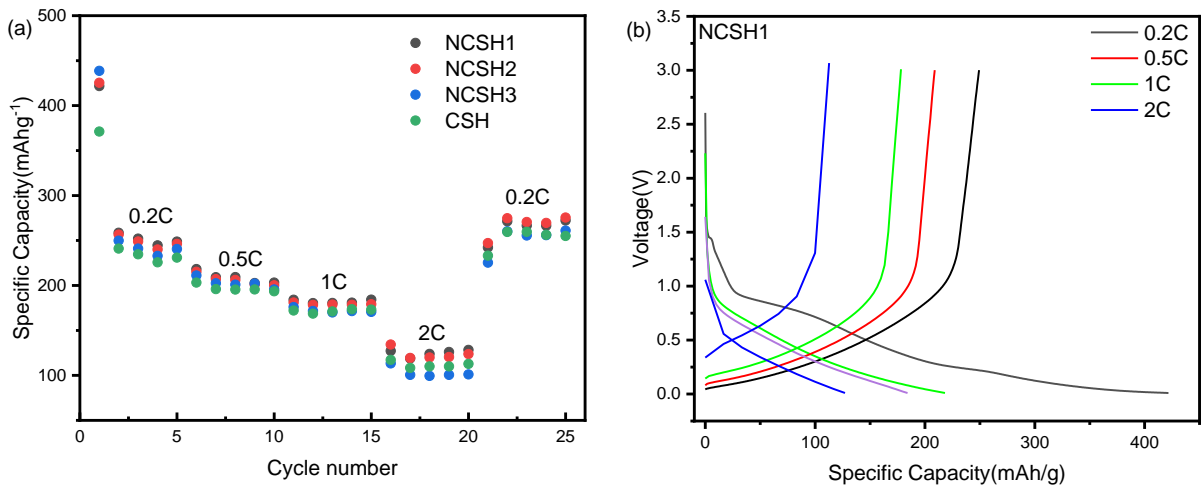

Figure 8. (a) Rate performance of different materials and (b) The first-week charge-discharge curves of NCSH1 at various magnifications

Cyclic voltammetry (CV) tests can be used to understand the chemical reaction situation of the electrode during the cycling process, and the test results are shown in Fig. 9. The reduction peak of NCSH1 is around 0.75 V, which corresponds to the formation peak of the solid electrolyte interphase (SEI) film in the first discharge process of the lithium-ion battery. The SEI film consumes lithium ions during the first cycle, which is one of the reasons for the low Coulombic efficiency of hard carbon materials. When the voltage is scanned from 0 to 2.5 V in the positive direction, it can be found that NCSH1 has a sharp oxidation peak at around 0.2 V, which represents the embed-ding peak of lithium ions in the material. There is a distinct hump around 1 V, which is due to the adsorption and desorption of lithium ions in the pores during the charging and discharging process, causing the formation of the hump. It can be found that the material has a high degree of overlap in the second and third cycles, indicating that the material has excellent cycle performance, which is consistent with the results of the cycle performance test of the material.

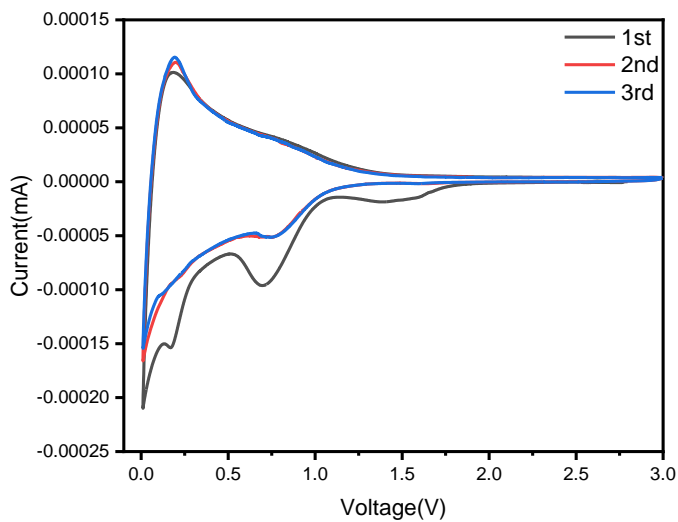

Figure 10. Cyclic voltammetry curves of NCSH1

Figure 11 shows the electrochemical impedance spectra (EIS) of NCSH1 and CSH. The low-frequency region of the AC impedance spectrum corresponds to the diffusion effect of lithium ions in the hard carbon negative electrode material, which is related to Warburg impedance. The semicircular arc in the high-frequency region is related to the charge transfer resistance and the interfacial impedance of the lithium-ion battery. It can be found that the curvature of NCSH1 in the high-frequency region is significantly increased compared to the undoped material, indicating that it has a smaller radius and lower impedance, which contributes to its better performance in the rate cycle. The results of the EIS circuit fitting are shown in Table 2. The results show that the transfer resistance and transmission resistance of NCSH1 are significantly reduced. Nitrogen elements, such as graphitic N and pyrrolic N, are doped into corn starch hard carbon materials in NCSH1, which reduces the impedance of the material, enhances its conductivity, and thus improves its rate performance.

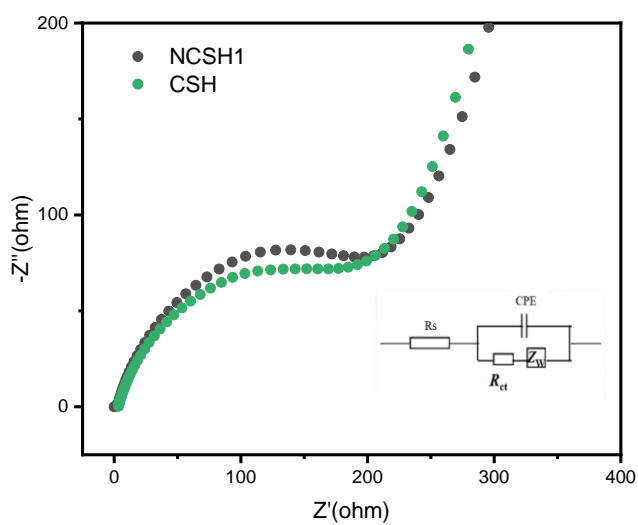

Figure 11. Electrochemical impedance spectra of NCSH1 and CSH-1000

## 3. Conclusions

The carbon materials obtained from corn starch can be used as anodes for lithium-ion batteries for reversible storage of lithium ions, with the storage capacity relative to the graphitization degree and surface defects. The interlayer spacing of the final hard carbon materials is greater than 0.38 nm, which is much larger than the interlayer spacing of graphite (0.34 nm) and has poor crystallinity and abundant defects, providing rich lithium storage sites as an anode for lithium-ion batteries, effectively promoting the rapid movement of lithium ions. During the presence of triethylenetetramine, the nitrogen atoms were uniformly doped into the corn starch hard carbon materials during the carbonization process, with nitrogen elements in the form of graphitic N, pyridinic N, and pyrrolic N embedded in the carbon materials; when triethylenetetramine and corn starch are carbonized at a mass ratio of 1:9, the nitrogen content of the obtained material is 0.91%, and the material exhibits good cycle performance, with a discharge capacity of 273.65 mAh g-1 after 100 cycles. The final results show that corn starch is one of the ideal precursors for hard carbon anode materials, and using triethylenetetramine to dope nitrogen into corn starch hard carbon is a simple and feasible method of nitrogen doping. Not only can it increase the discharge capacity of the material, but it can also reduce the transmission resistance of the material, effectively improving the overall electrochemical performance of the material.